\documentclass[prb,showpacs,
twocolumn,
aps]{revtex4}

\usepackage{graphicx}
\usepackage{dcolumn}
\usepackage{bm}

\begin{document}
\draft
\title{Electronic structure of the quasi-one-dimensional organic conductor TTF-TCNQ}

\author{M. Sing, U. Schwingenschl\"ogl, and R. Claessen}
\affiliation{Experimentalphysik II, Universit\"at Augsburg, D-86135 Augsburg, Germany}

\author{P. Blaha}
\affiliation{Institut f\"ur Physikalische und Theoretische Chemie, Technische
Universit\"at Wien, A-1060 Wien, Austria}

\author{J. M. P. Carmelo and L. M. Martelo\cite{Martelo}}
\affiliation{GCEP - Center of Physics, University of Minho, Campus
Gualtar, P-4710-057 Braga, Portugal}

\author{P. D. Sacramento}
\affiliation{Departamento de F\'{\i}sica and CFIF, Instituto Superior
T\'ecnico, P-1049-001 Lisboa, Portugal}

\author{M. Dressel}
\affiliation{1.~Physikalisches Institut, Universit{\"a}t Stuttgart, D-70550 Stuttgart, Germany}

\author{C. S. Jacobsen}
\affiliation{Department of Physics, Technical University of Denmark, DK-2800 Lyngby, Denmark}

\date{\today }

\begin{abstract}
We study the electronic structure of the quasi-one-dimensional
organic conductor TTF-TCNQ by means of density-functional band
theory, Hubbard model calculations, and angle-resolved
photoelectron spectroscopy (ARPES). The experimental spectra
reveal significant quantitative and qualitative discrepancies to
band theory. We demonstrate that the dispersive behavior as well
as the temperature-dependence of the spectra can be consistently
explained by the finite-energy physics of the one-dimensional
Hubbard model at metallic doping. The model description can even
be made quantitative, if one accounts for an enhanced hopping
integral at the surface, most likely caused by a relaxation of the
topmost molecular layer. Within this interpretation the ARPES data
provide spectroscopic evidence for the existence of spin-charge
separation on an energy scale of the conduction band width. The
failure of the one-dimensional Hubbard model for the {\it
low-energy} spectral behavior is attributed to interchain coupling
and the additional effect of electron-phonon interaction.
\\[1ex]
\end{abstract}


\maketitle

\section{Introduction}

The electronic structure of one-dimensional (1D) conductors provides a
valuable testing ground for the study of the quantum-mechanical
many-body problem. On the theoretical side there exist various
models for 1D interacting electron systems, which predict highly
unusual low-energy excitations due to dynamical decoupling of
charge and spin degrees of freedom. As a consequence, the {\it
low-energy} paradigmatic Fermi liquid picture fails for 1D metals
and a new generic many-body quantum state emerges which is
commonly referred to as Tomonaga-Luttinger liquid
(TLL).\cite{Voit95} Experimentally, quasi-1D metals are indeed
found to display marked deviations from conventional metallic
behavior, such as the absence of Fermi-Dirac edges in the
single-particle excitation spectra probed by angle-resolved
photoelectron spectroscopy (ARPES). \cite{Grioni00,Gweon01}
However, an unambiguous spectroscopic identification of
spin-charge separation and the existence of {\it low-energy} TLL
behavior in 1D metals is still lacking so far. Additional interest
in 1D electron systems arises from the suggestion that their
physics may also be relevant to the electronic structure of the
cuprate-based high-temperature superconductors,\cite{Orgad01}
related to the recent discovery of charge ordering in these
materials into narrow metallic 1D stripes separated by insulating
regions.\cite{Tranquada95,Tranquada98} Against this background the
search for {\it positive} spectroscopic signatures of unusual
electronic correlation effects in 1D metals remains to be of
topical importance.

In the search for promising realizations of a prototypical (quasi-)1D
conductor the organic charge transfer salts appear as interesting
candidates. Due to the formation of linear molecule stacks in the
crystal structure and an electronic charge transfer from cationic to anionic
complexes they display strongly anisotropic metallic
conductivities.\cite{Jerome82,Kagoshima88} Photoemission experiments
on such materials often find unusual spectral behavior like the
absence of a metallic Fermi edge. \cite{Grioni00} However, the lack of
information on surface quality, especially with regard to the rapid
photon-induced decomposition of organic compounds in the vacuum
ultraviolet,\cite{Sing03} casts serious doubts to what extent these
observations reflect intrinsic electronic properties or rather a
strongly disturbed surface. This is further corroborated by the
failure of ARPES to detect spectral energy-{\it vs.}-momentum
dispersions in most charge transfer
salts.\cite{Claessen02b,Soederholm97,Zwick97} A notable exception is
TTF-TCNQ (tetrathiafulvalene-tetracyanoquinodimethane), being the
first (and so far only) organic conductor for which dispersing 1D
bands have been observed by ARPES. \cite{Zwick98,Sing01} This
indicates a well-ordered periodic surface structure and thus lends
much enhanced significance to the observation of a deep pseudogap
around the Fermi energy, which even increases up to room temperature.
This spectral behavior has recently been interpreted as possible
indication of a highly unusual normal state in this 1D
conductor.\cite{Zwick98,Claessen02a}

In this paper we present a comprehensive experimental and theoretical
study of the electronic structure of TTF-TCNQ, elaborating on our
earlier ARPES results published in Ref.\onlinecite{Claessen02a}. The
comparison between experiment and band theory reveals significant
discrepancies, concerning both the width of the conduction bands as
well as their qualitative dispersion. While the band width
renormalization can be attributed to a molecular surface relaxation,
the remaining discrepancies indicate a failure of the bare band
picture. Rather, we are able to demonstrate that the TCNQ-derived part
of the ARPES finite-energy dispersions can be consistently mapped onto
the electron removal spectrum of the 1D Hubbard model at finite
doping.\cite{Carmelo03a,Carmelo03b} The importance of electronic
correlations is further corroborated by a peculiar temperature
dependence of the photoemission spectra. Based on these findings the
spectral behavior of TTF-TCNQ is interpreted as spectroscopic evidence
for spin-charge separation on an energy scale as large as the
conduction band width.

\section{Properties of TTF-TCNQ}

\begin{figure}
\includegraphics[width=16cm]{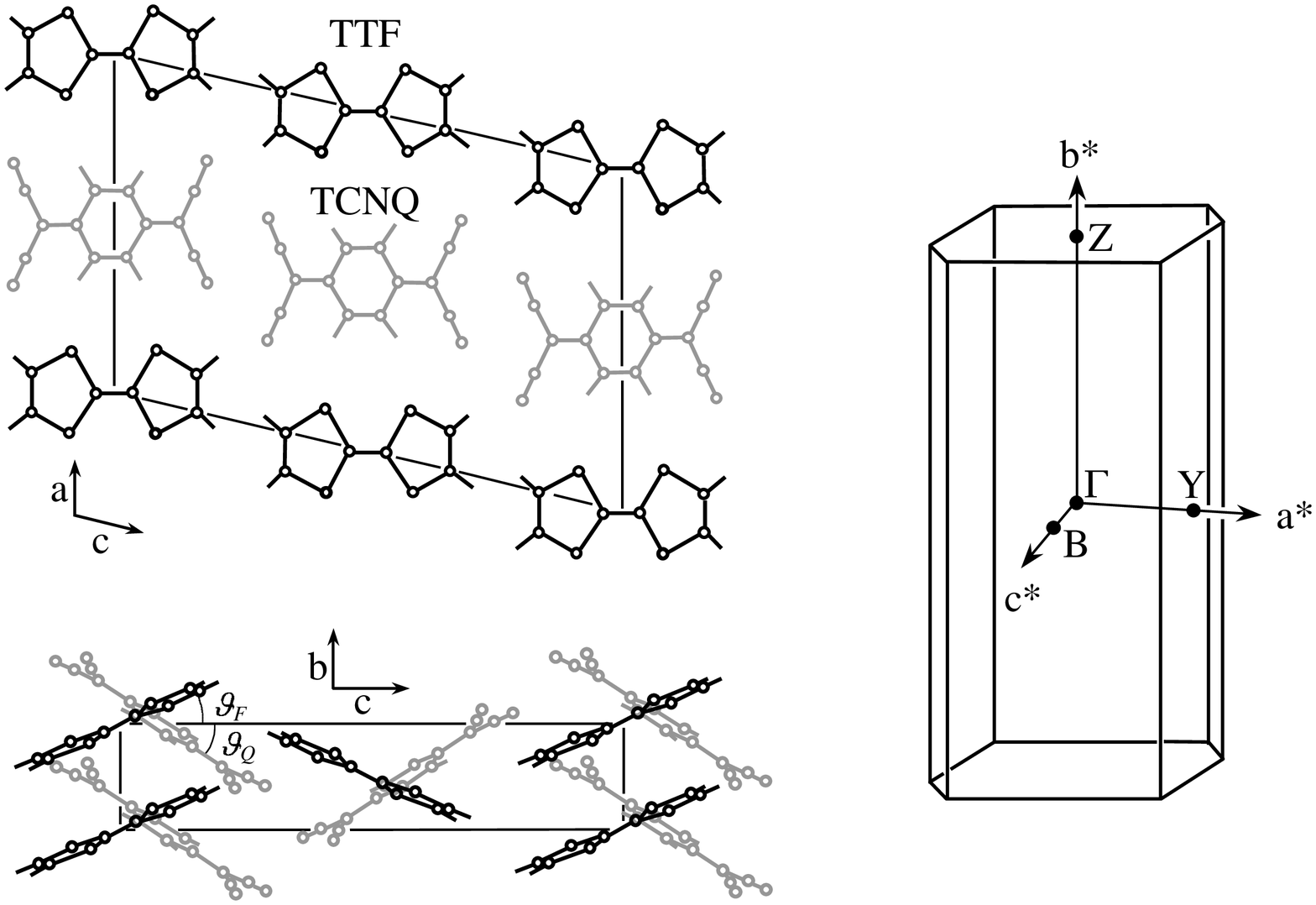}
\caption{Crystal structure of TTF-TCNQ. $\vartheta_F$ and
$\vartheta_Q$ indicate the tilt angles of the planar TTF and TCNQ
molecules, respectively, relative to the {\bf ac}-plane. Also
shown is the monoclinic Brillouin zone with its high symmetry
points.} \label{structure}
\end{figure}

The monoclinic crystal structure of TTF-TCNQ is shown in
Fig.~\ref{structure}. The lattice parameters at room temperature are
$a=12.298$ \AA, $b=3.819$ \AA, and $c=18.468$ \AA, the monoclinic
angle is $\beta = 104.46^{\circ}$. \cite{Kistenmacher74} The important
structural features are parallel linear stacks of planar TTF and TCNQ
molecules, respectively, oriented along the crystallographic {\bf b}
direction. The $\pi$-type molecular orbitals, extending over the
entire size of each molecule, overlap with those of the neighboring
molecules stacked above and below. Maximum covalent bonding is
achieved by tilting the molecular planes slightly about the {\bf
a}-axis, by $\vartheta_F = 24.5^{\circ}$ and $\vartheta_Q =
34.0^{\circ}$ for the TTF and TCNQ stacks, respectively
(cf.~Fig.~\ref{structure}). \cite{Kagoshima88,Kistenmacher74} The sign
of the tilt angle alternates between neighboring stacks, leading to
the herringbone structure of Fig.~\ref{structure}.

As covalent bonding occurs only along the stack direction, the
corresponding electronic TTF and TCNQ bands are expected to be
strongly anisotropic. Charge transfer of $\sim 0.59$ electrons per
molecule from TTF to TCNQ drives both types of chains metallic.
\cite{Kagoshima88,Pouget76} The conductivity along {\bf b} is two to
three orders of magnitude larger than perpendicular to it, making
TTF-TCNQ a truly quasi-1D metal. Below $T_P = 54$ K a charge density
wave (CDW) develops along the {\bf b}-direction, with wave-vector {\bf
Q}$_{CDW}=0.295$ {\bf b}* ($=0.485$ \AA$^{-1}$). The occurence of the
CDW is accompanied by a metal-insulator transition. From the activated
behavior of the conductivity a Peierls gap of $\sim 40$ meV has been
inferred. \cite{Kagoshima88} Within mean-field weak coupling theory
this translates into a transition temperature of $T_{MF}
\sim 125$ K. Due to the dominant effect of fluctuations in 1D systems
the actual Peierls transition is suppressed to about half of this
value. An additional transverse ordering transition occurs at 38 K.
\cite{Kagoshima88} The observation of diffuse x-ray scattering at $Q = 4k_F$
up to 220 K indicates the importance of electronic correlations.\cite{Pouget76}

\section{Band structure calculation}

The theoretical band structure was studied within the standard
density functional theory (DFT) approach using the generalized
gradient approximation (GGA).\cite{perdew96} We used the
self-consistent full-potential linearized augmented plane wave
(LAPW) method as implemented in the WIEN97 code. \cite{blaha97} A
basis set of about 12500 LAPWs and additional "local orbitals" for
the 2s (3s) states of C and N (S) were employed. This corresponds
to a lower basis set convergence than desirable, but was limited
by the available computational resources. Self-consistency was
achieved using 18 {\bf k}-points in the irreducible wedge of the BZ and
a temperature broadening scheme with 5 mRy. The results are
largely consistent with previous band calculations,
\cite{Berlinsky74,Kasowski92,Starikov98} but contain more detailed
and reliable information due to the more advanced method. Very
good agreement is found with the recent LDA/GGA pseudopotential
calculation of Ref.\onlinecite{Ishibashi00}. Our calculations have
been performed for the experimental room temperature
structure.\cite{Kistenmacher74} In addition, we also studied the
effect of structural distortions as model for a possible surface
relaxation (see section \ref{PES and DFT}).

\begin{figure}
\includegraphics[width=16cm]{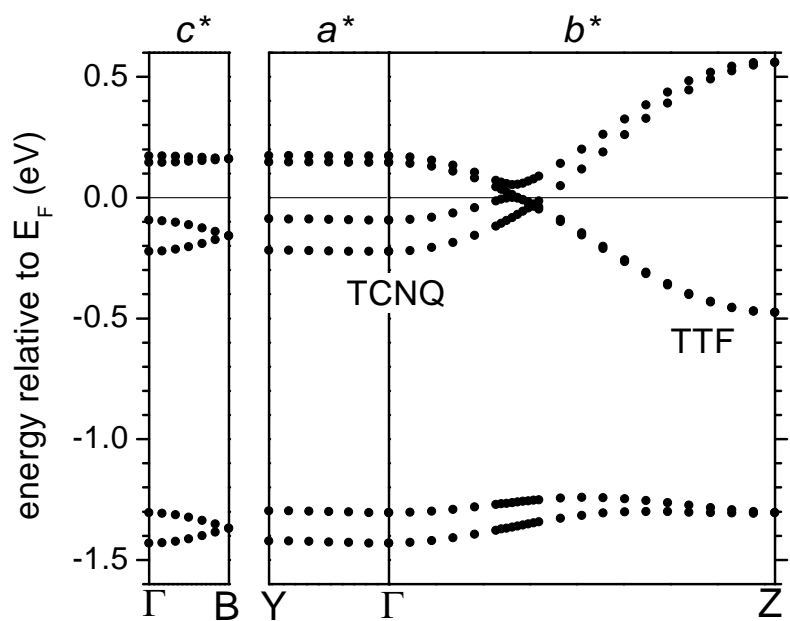}
\caption{DFT band structure near the Fermi level along the three
major high-symmetry lines of the Brillouin zone of TTF-TCNQ.}
\label{bandstructure}
\end{figure}

\begin{figure}
\includegraphics[width=16cm]{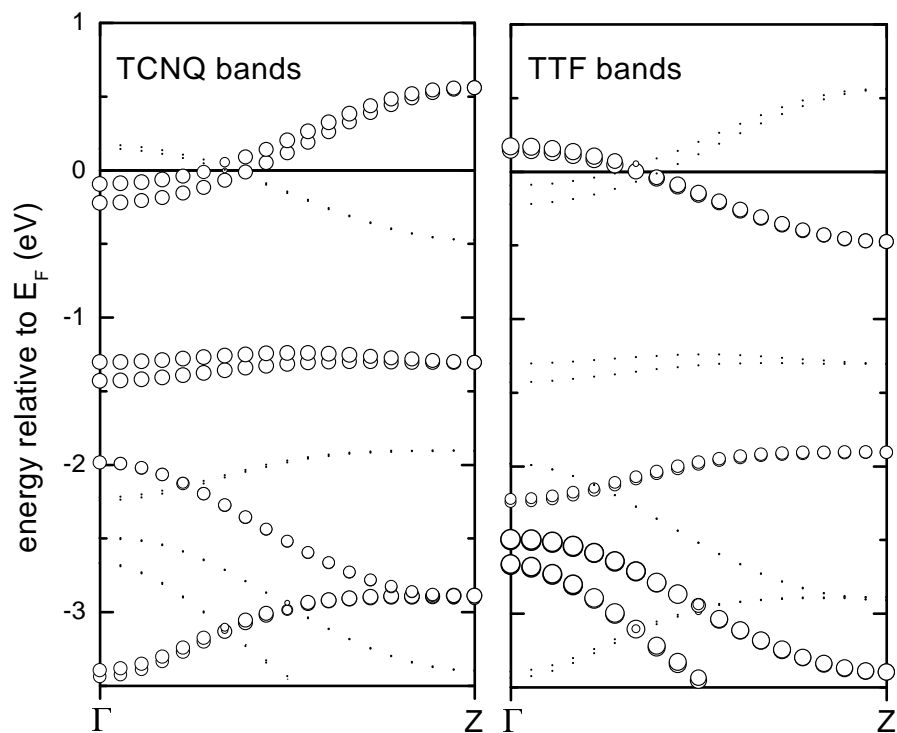}
\caption{Theoretical band dispersions along $\Gamma$Z showing the
molecular origin of the bands. The size of the symbols represents
the charge of each state residing on the TCNQ (left panel) and TTF
(right panel) molecules.} \label{band origin}
\end{figure}

According to the DFT calculation TTF-TCNQ is characterized by
strong intramolecular covalent bonding, whereas the interaction
between the molecules is predominantly ionic. Thus, the molecular
orbitals are strongly localized except along the stacking
direction, where small but notable covalent intermolecular bonding
occurs. As seen in Fig.~\ref{bandstructure}, this leads to the
formation of two sets of quasi-1D conduction band doublets with
pronounced dispersion along {\bf b}*, {\it i.e.} the $\Gamma$Z
line of the Brillouin zone (cf.~Fig.~\ref{structure}). The first
one, just below $E_F$ at the $\Gamma$-point and unoccupied at Z,
is derived from $\pi$-bonded 2p orbitals of mostly C(6) atoms
(in the notation of Ref.~\onlinecite{Kistenmacher74}) and can thus
be attributed to the TCNQ stacks. The 130 meV splitting at
$\Gamma$ results from a weak interaction between the two TCNQ
stacks in the unit cell. The other conduction band doublet,
showing the opposite dispersion and remaining nearly degenerate,
is mainly derived from the $3 {\rm p}_{\pi}$ orbitals of S(1) and
S(2) atoms and therefore associated with the TTF stacks. A small
hybridization gap opens between the respective upper TCNQ and TTF
bands at the Fermi level. The definite assignment of each band to
either the TCNQ or the TTF stacks can also be seen in
Fig.~\ref{band origin}, which for each band state shows the
electronic charge localized on the TCNQ and the TTF molecules,
respectively.

The metallic nature of the TTF-TCNQ band structure arises from the
energetic overlap of the quasi-1D TCNQ and TTF bands and the
electronic charge transfer between them. Due to interstack
interaction our calculation yields two Fermi vectors, $k_F=0.27$
and $0.33$ \AA$^{-1}$, slightly larger than expected from the
nesting vector $2k_F=0.485$ \AA$^{-1}$ derived from the CDW
periodicity. The theoretical bandwidths along $\Gamma$Z are 0.7 eV
(TCNQ) and 0.65 eV (TTF), in fair agreement with experimental
estimates of $\sim$0.5 eV. \cite{Kagoshima88}

The conduction band dispersion perpendicular to {\bf b}* is
essentially negligible. Along the $\Gamma$Y line ({\bf a}*) it is
practically zero. Slight dispersion of the TCNQ-derived band
occurs along $\Gamma$B, which results from weak interaction along
{\bf c} between the molecular end groups in neighboring TCNQ
stacks. However, the effect is too small to cause a band crossing
along $\Gamma$B. The resulting Fermi surface topology is therefore
truly 1D.

\section{Angle-resolved photoemission}

ARPES measurements have been performed at our home lab, using He I
radiation from a discharge lamp and an Omicron EA 125 HR electron
energy analyzer, and with synchrotron radiation at BESSY (Berlin)
using an Omicron AR 65 spectrometer.\cite{Janowitz99} In both cases
the energy and angular resolution amounted to 60 meV and $\pm
1^{\circ}$, respectively. All data were taken above the Peierls
transition at a sample temperature of 60 K. TTF-TCNQ single crystals
were grown by diffusion in pure acetonitrile and had typical dimensions of $2
\times 5 \times 0.2$ mm$^3$, with the long axis along {\bf b}. Their
quality was characterized by x-ray diffraction, electron spin
resonance, and conductivity measurements. Clean surfaces parallel to
the {\bf ab}-plane were obtained by {\it in situ} cleavage of the
crystals at a base pressure of $< 10^{-10}$ mbar. The stoichiometry of
this surface, which contains both TCNQ and TTF chains,
\cite{Kistenmacher74} was verified by x-ray photoemission.
\cite{Sing03} From the observation of momentum-dispersive ARPES
structures we conclude on a crystalline long-range order of the
surface, which has independently been confirmed by scanning tunneling
microscopy (STM).\cite{Sleator88,Wang03}

\begin{figure}
\includegraphics[width=16cm]{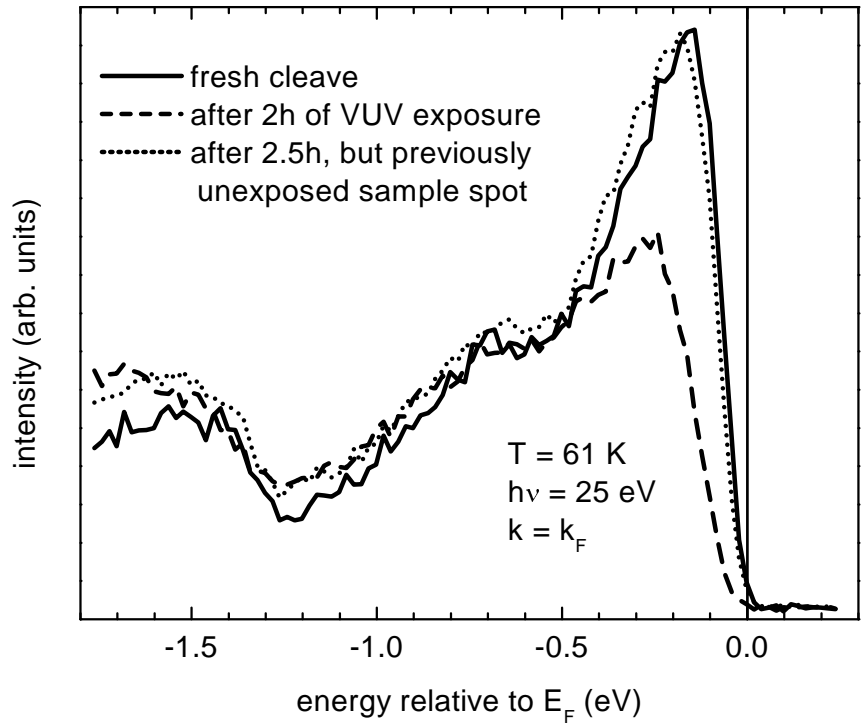}
\caption{Angle-resolved energy distribution curves at the Fermi
vector showing the effect of photon-induced surface degradation
($h \nu = 25$ eV, $T = 61$ K). For a detailed discussion see
text.} \label{photon effects}
\end{figure}

Great care was taken to avoid photon-induced surface damage by
minimizing the exposure to the incident 25 eV radiation. The effect is
demonstrated in Fig.~\ref{photon effects}, which contains spectra
taken at the experimental Fermi vector. For a freshly prepared
surface, {\it i.e.} immediately after cleavage of the crystal, an
intense peak is observed close to the Fermi level. After two hours of
exposure to vacuum ultraviolet (VUV) radiation its intensity has
strongly decreased and its peak position shifted by more than 0.1 eV
away from $E_F$. However, the original spectrum can be completely
recovered by taking data on another previously unexposed sample spot.
This demonstrates that the observed surface degradation is not simply
due to contamination or decomposition in the vacuum but indeed caused
by VUV radiation. Unfiltered higher order light or direct use of
higher photon energies ($^>_{\sim} $35 eV) reduces the time scale of
the VUV-induced surface damage even down to minutes. \cite{Sing03} All
data presented in the remainder of this paper were obtained before
noticeable surface decomposition occured.

\begin{figure}
\includegraphics[width=16cm]{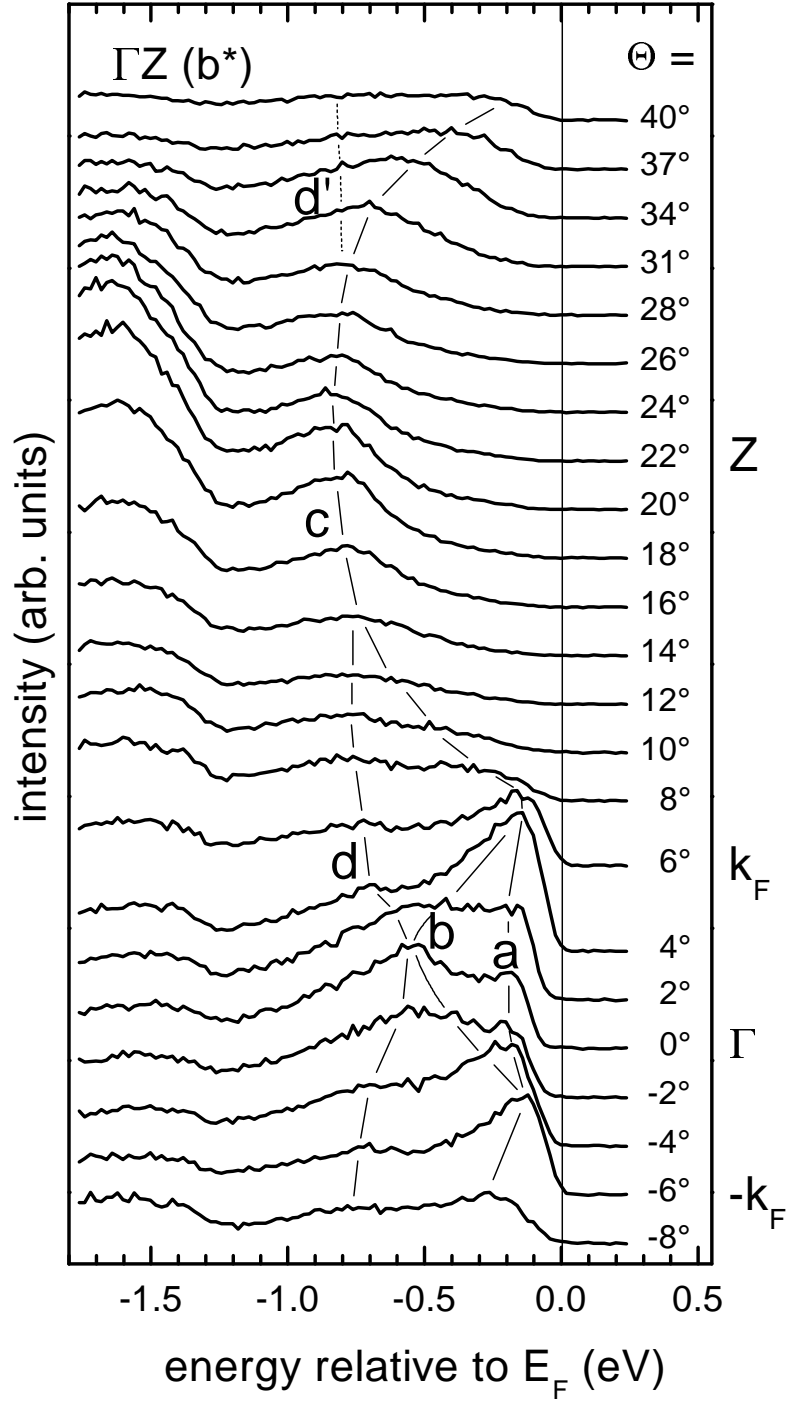}
\caption{Energy distribution curves measured along the $\Gamma$Z
direction ($h \nu = 25$ eV, $T = 61$ K). The thin lines are guides
to the eye and are meant to indicate the dispersion of the
spectral features.} \label{EDCs}
\end{figure}

Figure \ref{EDCs} shows energy distribution curves obtained along the
${\bf b^*}$ axis, {\it i.e.} the 1D direction. The spectral features
display pronounced dispersion, whereas spectra measured perpendicular
to ${\bf b^*}$ are dispersionless (not shown here, see
Refs.\onlinecite{Sing03,Zwick98}). Our data are in excellent agreement
with those of Zwick {\it et al.}\cite{Zwick98} but display in parts
more spectral detail. For example, in normal emission ($\theta =
0^{\circ}$) we can clearly distinguish two peaks at 0.19 and 0.54 eV
below $E_F$, labeled {\textsf a} and {\textsf b} in Fig.~\ref{EDCs}.
In Ref.~\onlinecite{Zwick98} peak {\textsf a} appeared only as a
shoulder and was not discussed. For off-normal emission peak {\textsf
b} splits into two parts. The upper one (retaining the label {\textsf
b}) moves upwards in energy and converges with {\textsf a} close to
$\theta = 6^{\circ}$, where both features reach their closest approach
to the Fermi level. We identify this position as Fermi vector which
yields $k_F = 0.24 \pm 0.03$ \AA$^{-1}$, in good agreement with the
value derived from the CDW vector. Note however, that despite the high
conductivity no metallic Fermi edge is observed in the $k_F$-spectra
(see also Fig.~\ref{photon effects}), as already observed in
Ref.~\onlinecite{Zwick98}. The spectral intensity rather decreases
almost linearly down to zero at exactly the Fermi level.

Beyond $k_F$ a weak structure {\textsf c} moves back again from
the Fermi level and displays a dispersion symmetric about $\theta
= 22^{\circ}$ corresponding to the Z-point of the Brillouin zone.
\cite{criticalpoints} Returning to the splitting of peak {\textsf
b} away from $\theta = 0^{\circ}$ we note that its lower part
(labeled {\textsf d}) disperses downwards in energy and
eventually becomes obscured by peak {\textsf c}. For very high
emission angles, corresponding to a $\bf k$-vector in the next
zone, one observes a symmetry-related weak shoulder {\textsf
d$^\prime$}.

\section{Temperature dependence of the photoemission spectra}
\label{T-dependence}

\begin{figure}
\includegraphics[width=16cm]{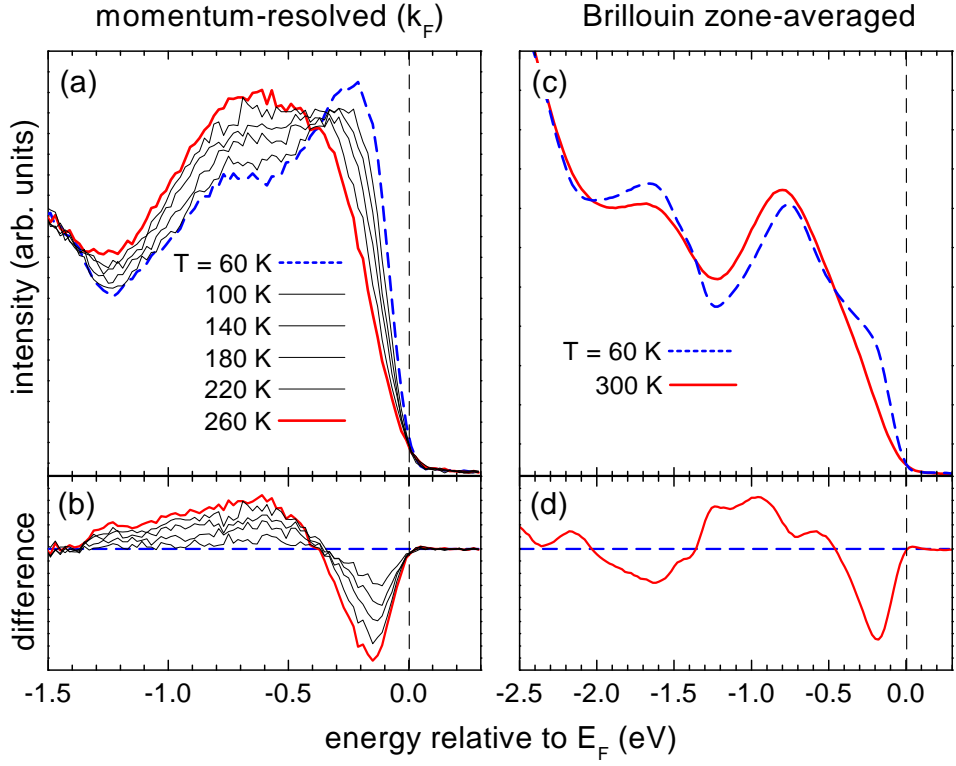}
\caption{Temperature dependence of the photoemission spectra
($h\nu = 21.2$ eV): (a) Momentum-resolved spectrum at $k=k_F$ measured
between 60 K (blue/dashed curve) and 260 K (red/solid curve). (b)
Difference spectra relative to 60 K. (c) Momentum-integrated spectrum
at 60 K (blue/dashed) and 300 K (red/solid); note the larger energy
scale compared to (a) and (b). (d) Difference spectrum relative to 60
K.} \label{T-spectra}
\end{figure}

Upon cooling through the Peierls temperature the Fermi vector spectrum
has been shown to display the expected opening of a CDW
gap.\cite{Zwick98} However, even more remarkable is the temperature
dependence of the $k_F$-spectrum {\it above} the
transition.\cite{Zwick98,Claessen02a} This is shown in
Fig.~\ref{T-spectra}(a) where considerable spectral changes are
observed between 60 and 260 K. The interpretation of these changes
hinges on a careful intensity normalization of the spectra. This has
been achieved by normalizing them on the residual background intensity
above the Fermi energy, which is a good measure of the exciting photon
flux.\cite{footnote_normalization} Incidentally, this procedure leads
to an almost complete alignment of the temperature spectra at high
binding energies ($^{>}_{\sim} 1.3$ eV). At lower energies the
spectral changes from low to high temperatures can then be described
as a pronounced intensity loss of the peak near the Fermi level (and
its slight shift away from $E_F$), while at the same time the
intensity increases between -0.4 and -1.3 eV (see difference spectra
in Fig.~\ref{T-spectra}(b)). With the described normalization the {\it
integrated} spectral weight remains however conserved within
experimental uncertainty. We also note that these temperature effects
are fully reversible.

These observations and, in particular, the conservation of the total
intensity suggest that at $k=k_F$ spectral weight is transferred from
low to high binding energies with increasing temperature. However,
based on the temperature dependence of the $k_F$-spectrum alone we
cannot rule out the possibility that the effect is caused by a
redistribution of spectral weight in {\it momentum space} rather than
in energy, caused, e.g., by phonon-induced disorder which may be large
in organic compounds. In order to check this we have determined the
$k$-integrated density of states (DOS) by summing up ARPES spectra
covering the entire 1D Brillouin zone from $\Gamma$ to Z.
Fig.~\ref{T-spectra}(c) shows the result for $T= 60$ and 300 K. A
smearing of the spectral weight distribution in momentum space due to
thermally excited phonons should have no effect on the $k$-integrated
energy spectrum, except possibly for the phonon-induced lifetime
broadening of the spectral peaks. The latter seems to be the case for
the temperature change around $-1.6$ eV (the corresponding ARPES
spectra show that the broadening occurs only near the zone edge,
indicating a particularly strong electron-phonon coupling there). At
low binding energies we recover the temperature dependence of the
$k_F$ spectrum (cf.~the difference spectra in Fig.~\ref{T-spectra}(b)
and \ref{T-spectra}(d)), which can clearly not be explained by line
broadening. We hence conclude that the temperature behavior at the
Fermi vector is indeed caused by a spectral weight transfer in energy.

\section{Comparison of photoemission and band theory}
\label{PES and DFT}

The identification of the ARPES dispersions indicated by the thin
lines in Fig.~\ref{EDCs} is further substantiated by a different
representation of the data. Figure~\ref{intensity map} shows the
negative second energy derivative of the photocurrent
$-d^2I/dE^2$, clipped at zero value, as grayscale plot in the
$(E,k)$-plane. This enhances the visibility of the spectral
structures and visualizes their dispersion in a completely
unbiased way. Also shown are the DFT conduction bands. The
comparison of experiment and theory reveals qualitative
similarities but also significant discrepancies. Starting our
discussion with experimental structure {\textsf c} we find its
dispersive behavior in agreement with that of the theoretical
TTF-derived bands, except that the (occupied) bandwidth exceeds
the theoretical one by a factor of $\sim 1.7$. Similarly,
structures {\textsf a} and {\textsf b} can be attributed to the
theoretical TCNQ doublet bands if one accounts for largely enhanced
({\textsf a}: $\sim 2.0$ and {\textsf b}: $\sim 2.4$) band widths.
Finally, we point out that experimental feature {\textsf d} finds
{\it no} counterpart in the band calculation.

\begin{figure}
\includegraphics[width=16cm]{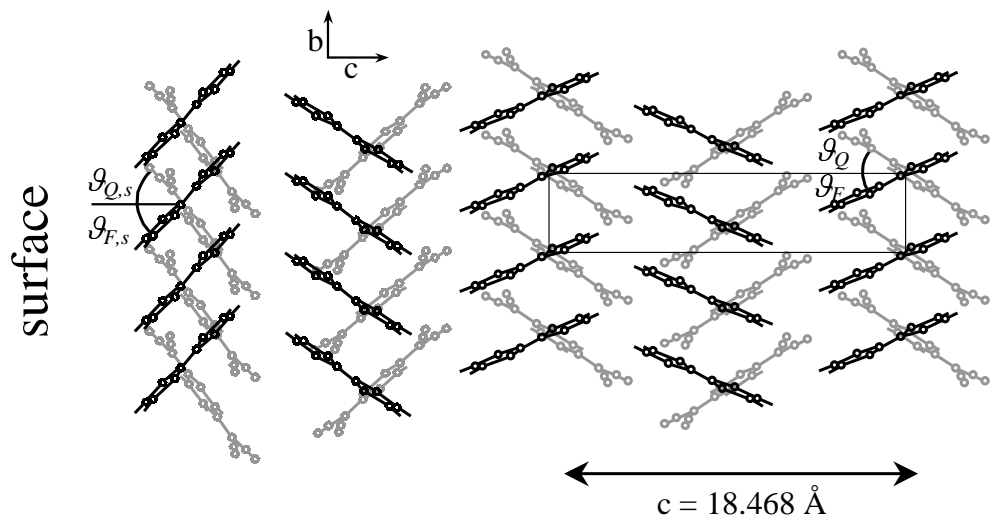}
\caption{Schematic picture of a possible surface relaxation
leading to enhanced molecular tilt angles $\vartheta_{Q,s}$ and
$\vartheta_{F,s}$ of the topmost TCNQ and TTF molecules,
respectively.} \label{surface relaxation}
\end{figure}

Our experimental conduction band widths are not only at variance
with band theory but also clearly exceed the estimates derived
from bulk-sensitive measurements. \cite{Kagoshima88} As the ARPES
probing depth is comparable to the thickness of a single molecular
layer ($c/2=9.23$ \AA, cf.~Fig.~\ref{surface relaxation}),
the observed discrepancies suggest that the electronic structure
of the topmost layer(s) differs from that of the volume. One
possible origin could be a structural surface relaxation involving
the tilt angles of the planar (and relatively rigid) TTF and TCNQ
molecules, respectively, relative to the {\bf b}-axis. We note
that these angles correspond to a total energy minimum
configuration resulting from a competition between maximum
covalent bonding along the stack direction and minimum Coulomb
energy in the Madelung potential of the surrounding molecular
ions.\cite{Phillips73} It seems conceivable that at the surface
this balance is offset due the altered Madelung potential, leading
to different equilibrium tilt angles of the topmost molecules, as
sketched in Fig.~\ref{surface relaxation}. If the surface tilt
happens to be larger than that in the volume, it will result in a
reduced separation between the molecular planes within a given
stack. This in turn leads to an increase of the intermolecular
hopping integral and hence the bandwidth.

We have tested this idea by performing a band calculation for a
hypothetical volume structure with increased tilt angles
(approximately doubled relative to their bulk values), which indeed
leads to a strongly enhanced band width at least for the TTF-derived
bands in good agreement with their ARPES dispersion. However, a
realistic calculation of the surface relaxation by total energy
optimization for a semi-infinite crystal is currently beyond our
technical limits owing to the large size of the TTF-TCNQ unit cell.
Unfortunately, a reliable {\it experimental} determination of the
molecular surface tilt seems also out of reach, as the usual methods
for surface structure determination do not work here. Low energy
electron diffraction (LEED) of TTF-TCNQ is strongly hampered by
electron-induced surface damage even faster than that caused by the
VUV photons.\cite{Sing03} STM as another important structural surface
probe is only capable of determining the surface periodicity
\cite{Sleator88,Wang03} but cannot give any reliable information on
molecular off-plane orientation. Therefore, the suggested surface
relaxation has to remain a mere speculation at this point. However,
whatever its microscopic origin, the observed enhancement of the ARPES
bandwidth with respect to the volume is an experimental fact and we
thus have to accept it as an established property of the probed
surface layer which distinguishes it from the bulk.

We are finally left with feature {\textsf d}, which even under the
assumption of a surface band width renormalization cannot be
identified with any of the theoretical volume bands. It might
appear tempting to attribute it to an intrinsic surface state.
However, such an interpretation is in conflict with the observed
Fermi vector of the other bands: Since {\textsf d} stays well
below the Fermi level and would thus be occupied throughout the
entire Brillouin zone (in a one-electron band picture), it should
severely affect the delicate charge balance between the TTF and
TCNQ bands and shift the surface Fermi vector notably from its
bulk value, which is not the case. Another explanation of {\textsf
d} as backfolded image of the TTF band induced by long-ranged CDW
fluctuations \cite{Schaefer01} is ruled out due to the lack of
other evidence for backfolding in the data. As we will discuss in
the following section, spectral feature {\textsf d} finds a
natural explanation as a many-body effect.

\section{Comparison to the 1D Hubbard model}

There is substantial experimental evidence that Coulomb
interaction plays an essential role in the electronic structure of
TTF-TCNQ \cite{Jerome82,Kagoshima88,Torrance77,Basista90} and that
a purely band theoretical description may be inadequate. On the
theoretical side, the dramatic effects of electron-electron
interaction on the {\it low-energy} properties of 1D metals have
been studied in much detail using the Tomonaga-Luttinger (TL)
model. \cite{Voit95} It is based on a 1D conduction band with
infinite linear dispersion and treats interaction by including
scattering processes about the Fermi points.\cite{Voit95,Meden92}
The TL model focusses on the low-energy physics and describes in
detail the breakdown of the quasiparticle picture for the
low-lying excitations and the emergence of TLL behavior resulting
from the dynamical decoupling of spin and charge. For example, the
TL model predicts a low-energy onset of the single-particle
spectrum which is no longer given by a metallic Fermi edge but
rather by a power law behavior $\propto \omega^{\alpha}$, with the
exponent $\alpha$ determined by the coupling parameters of the
model. The low-energy physics of the TL model defines in fact a
universality class which includes also more complicated 1D models
of interacting electrons.\cite{Solyom79,Haldane81} However, by its
very construction the TL model contains no intrinsic energy scale,
and therefore the energy range of its applicability to real 1D
metals is principally unknown, making it less useful for the study
of {\it finite-energy} spectral properties.

As seen in the previous section, the ARPES data of TTF-TCNQ indeed
show unusual behavior on an energy scale of the {\it entire band
width}, not just near the Fermi level. The spectral properties
over this much wider energy range have so far only been addressed
by the 1D single-band Hubbard model.
Compared to band theory it appears as a much better starting point
for the description of TTF-TCNQ and other organic charge transfer
salts. In fact, various properties of these 1D conductors have
already successfully been analysed within a Hubbard model
framework, such as the magnetic susceptibility
\cite{Torrance77,Lee77} or the nuclear spin relaxation
rate.\cite{Jerome82,Devreux76} The underlying idea is that the
local interaction energy $U$ for two electrons residing on the
same molecule dominates over long range Coulomb contributions. The
delocalization of the charge carriers is described by the hopping
integral $t$ (the bare bandwidth amounts to $4t$ in one
dimension). The Hubbard model also defines an intrinsic energy
scale for spin excitations, which for large values of $U/t$
is given by the exchange constant $J={{2t^2}\over U}\,\Bigl[n -
{\sin(2\pi n)\over 2\pi}\Bigr]$, \cite{Coll74,Klein74} with $n$
being the band filling parameter ($n = 0.59$ for TTF-TCNQ).

At low excitation energies the physics of the 1D Hubbard model with
finite doping follows the TLL phenomenology. The $U/t$-dependence of the
non-classic TLL exponents, which control the asymptotics of the
low-energy correlation functions, can be extracted from its
Bethe-ansatz solution.\cite{Lieb68} However, in contrast to the TL
model the Hubbard model also allows the study of finite-energy
properties. Recently, an exact analysis on the basis of the
Bethe-ansatz has shown \cite{Carmelo03a,Carmelo03b} that all energy
eigenstates of the 1D Hubbard model can be described in terms of
occupancy configurations of various collective spin-only and
charge-only modes, namely {\it spinons} (zero-charge spin
excitations), {\it holons} (spinless charge excitations), and a third
type of charged quantum objects. \cite{footnote_terminology} We refer
the reader to Refs.~\onlinecite{Carmelo03a} and
\onlinecite{Carmelo03b} for details. The important point is that this
description is valid for {\it all} energy scales of the model and
follows from the non-perturbative organization of the electronic
degrees of freedom.

Here we are interested in the electron removal spectrum of the 1D
Hubbard model. Qualitative properties of the spectrum have already
been derived from early calculations within the strong coupling limit
($U/t\rightarrow\infty$).\cite{Penc96,Favand97} More recently, it has
become possible to determine the spectral behavior also for
intermediate interaction strengths ($U \sim 4t$), either by numerical
methods\cite{Senechal00} or by
Bethe-ansatz.\cite{Carmelo03a,Carmelo03b} Here we will restrict
ourselves to the energy-{\it vs.}-momentum dependence of the spectral
features, which can be calculated exactly with the latter method. The
calculation of matrix elements between ground and excited states and
hence of the spectral weight distribution is more complicated with
this method and will be presented elsewhere.\cite{Carmelo03c} A
schematic picture of the spectral dispersions is given in
Fig.~\ref{Hubbard model spectra}(a). As the hole generated by the
removal of one electron decomposes (or ''fractionalizes'') into
decoupled spin and charge excitations, there is a manifold of ways to
distribute excitation energy and momentum among these collective modes
giving rise to an excitation continuum, indicated by the shaded area
in Fig.~\ref{Hubbard model spectra}(a).

However, due to the phase space available for electronic hole
fractionalization this continuum is not structureless. It is dominated
by lines of singularities (solid curves in Fig.~\ref{Hubbard model
spectra}(a)) which roughly speaking correspond to situations, in which
either the charge mode propagates with the entire excitation energy
leaving zero energy for the spin channel, or {\it vice versa}. We
denote these dispersion curves hence as ''charge'' and ''spin''
branches, respectively. At the Fermi vector both branches are
degenerate, but due to their different group velocities they split
away from the Fermi level. This low-energy behavior has already been
found for the TL model.\cite{Voit95} The 1D Hubbard model now allows
us to explore also the finite-energy dispersion of these features. The
spin branch for example reaches its maximum binding energy for
momentum $k=0$ at about $\frac{\pi}{2} J$, reflecting the dispersion
of a bare {\it spinon}. The dispersion of the charge part is a little
bit complicated. Starting from $k=-k_F$ it follows a nearly
parabolic-like dispersion reaching at $k = +k_F$ a maximum binding energy
which scales with the hopping integral $t$. From there it disperses
upwards again under a pronounced loss of spectral weight
\cite{Penc96,Favand97,Senechal00} until it eventually crosses the Fermi level
at $+3k_F$ (for symmetry reasons there is a corresponding charge
branch running from $+k_F$ to $-3k_F$). The shape of its dispersion
reflects that of a bare {\it holon}, with the distance of $4k_F$
(rather than $2k_F$) between its Fermi level crossing points owing to
the fact that the {\it holon} is a spinless quantum object. The shift
of the symmetry point away from $k=0$ to $\pm k_F$ can be understood
from a detailed microscopic analysis of the electronic hole
spectrum.\cite{Carmelo03a,Carmelo03b} The peculiar high-energy
behavior of the charge branch was first noted by Penc {\it et al.} in
the case $U/t \rightarrow \infty$.\cite{Penc96}

\begin{figure}
\includegraphics[width=16cm]{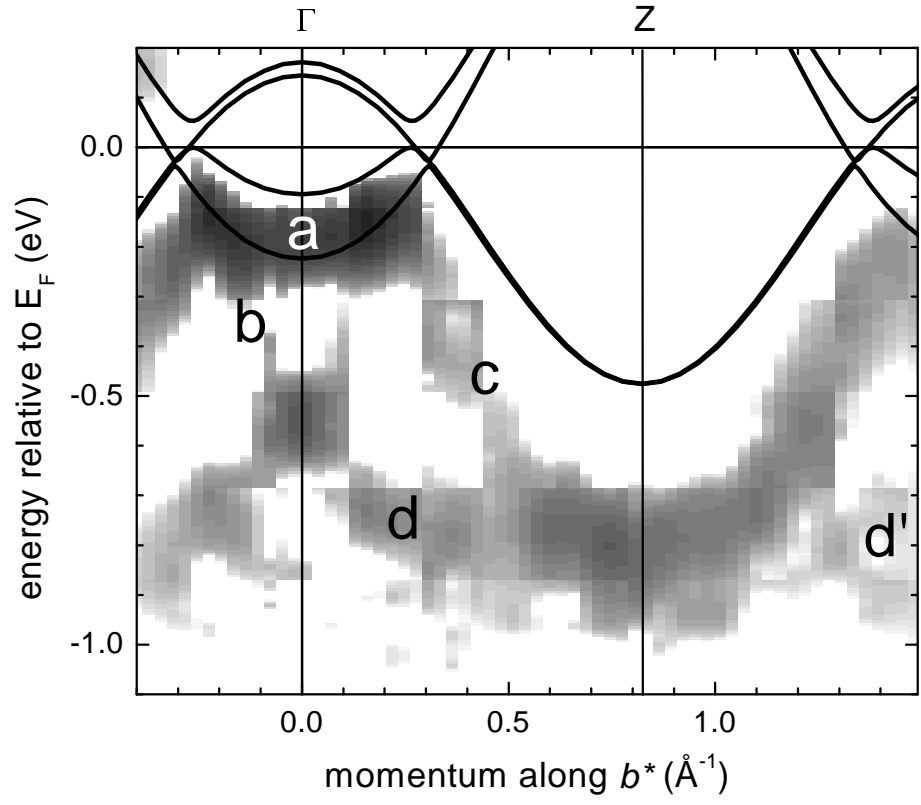}
\caption{Gray-scale plot of the ARPES dispersions (see text for
details) in comparison to the conduction band dispersions obtained
by density functional theory.} \label{intensity map}
\end{figure}

Comparing this picture to the observed experimental dispersions of the
TCNQ-related peaks in the ARPES data (cf.~Fig.~\ref{intensity map}) we
find remarkable similarities. In fact, it is even possible to obtain a
{\it quantitative} Hubbard model description of the experimental
dispersions. For this purpose we have utilized the Bethe-ansatz method
introduced in Ref.~\onlinecite{Carmelo03c}. Further details about the
line shape predicted by the Hubbard model within such a method for the
TCNQ dispersions are presented elsewhere.\cite{Carmelo03d} The method
leads to $U/t$ dependent branch lines which are given by the
expressions of Ref.~\onlinecite{Carmelo03a} [Eqs. (C15), (C16), (C19),
and (C21)] for the bare {\it holon} and {\it spinon} dispersions,
\cite{footnote_terminology} which in turn reproduce those of the
charge and spin branches in the electron removal spectrum. For these
calculations the model parameters $U$ and $t$ were chosen in such a
way as to yield optimum agreement with the ARPES dispersions. The
comparison of the model calculation to the experimental TCNQ
dispersions (from ARPES spectra measured on a finer $k$-grid than
those presented above) in Fig. \ref{Hubbard model spectra}(b) yields
an almost perfect match. This allows us to identify the experimental
structures {\textsf a} and {\textsf b} (cf.~Fig.~\ref{intensity map})
as spin branch and the upper part of the charge branch, respectively.
\cite{footnote_on_splitting} Moreover, the as yet unidentified
structure {\textsf d} now finds its natural explanation as the
high-energy part of the theoretical charge branch, at least for not
too large $k$-vectors. Experimentally, its reversed dispersion beyond
$k_F$ and its eventual $3k_F$-crossing is not observed, most likely
due to the theoretically predicted loss of weight at larger $k$ and
the overlapping TTF band. The model parameters used to fit the
theoretical dispersions to the experimental ones are $t=0.4$ eV and
$U=1.96$ eV, corresponding to a rather moderate coupling strength of
$U/t = 4.9$.

We finally turn to the TTF-related ARPES feature. Concerning its
dispersion we observe no extraordinary behavior other than the
enhanced bandwidth relative to band theory. Complimentary to the TCNQ
band, the TTF-derived conduction band (or rather band doublet) is more
than half-filled ($n= 2-0.59 > 1$). For this case the Hubbard model
predicts a charge branch line whose dispersion shows some similarities
to that of the experimental feature {\textsf c} in Fig.~\ref{intensity
map}. However, there should also be an additional spin branch line for
which we observe no clear evidence. It is possible that the model
parameters suitable for the TTF band differ from those for the TCNQ
band and are such that spin-charge separation is less pronounced.
Moreover, x-ray diffraction studies have reported the existence of
$4k_F$ CDW fluctuations on the TTF chains up to 220
K.\cite{Kagoshima88,Pouget76} It is not clear how such fluctuations
may affect the ARPES spectra; they could for example
account for the relatively large line-width of the experimental TTF
peak (cf.~structure {\textsf c} in Fig.~\ref{EDCs}), thereby obscuring
a possible small {\it spinon}-{\it holon} splitting. At this point the
interpretation of the TTF part of the ARPES data has to remain an open
question.

\section{Discussion}

The 1D Hubbard model thus provides a quantitative description of the
experimental TCNQ-related dispersions and an explanation for the
failure of band theory. In fact, earlier studies of the electronic and
magnetic low-energy properties of TTF-TCNQ
\cite{Jerome82,Kagoshima88,Torrance77,Basista90} have already used
this model successfully for the interpretation of their data. They
estimated that the local interaction energy $U$ and the bandwidth $4t$
are comparable and of the order of 1 eV. This is consistent with the
parameters of our model fit. Concerning the resulting bandwidth $4t =
1.6$ eV we observe an approximate doubling with respect to the result
of our DFT calculation (0.7 eV), just as in the bare band-theoretical
interpretation of the ARPES data. It is again attributed to a possible
molecular surface relaxation as already discussed in Section \ref{PES
and DFT}. The hopping integral of our Hubbard model fit thus reflects
a surface property. In order to compare to $bulk$ properties we should
rather use the value inferred from the DFT bandwidth, $t=0.175$ eV. As
the intramolecular Coulomb energy $U$ is a local quantity, we do not
expect large differences between bulk and surface. With $U = 1.96$ eV
we thus obtain for the coupling strength in the volume a value of $U/t
= 11.2$. From $U$ and $t$ we can also calculate the magnetic exchange
energy $J$, which for the volume yields 21 meV (110 meV for the
surface). This is in good agreement with experimental estimates of $J$
inferred from magnetic susceptibility measurements which range between
$17$ and $30$ meV.\cite{Torrance77,Klotz88}.

Further evidence for the importance of correlation effects is
provided by the unusual temperature dependence of the
photoemission spectra. Commonly, temperature effects are caused by
electron-phonon interaction with spectral changes occuring on an
energy scale $k_B T$, due to an altered population of phonons with
a comparable energy (at least within harmonic approximation and
with linear coupling).\cite{Knupfer93} However, for TTF-TCNQ we
observe upon warming-up a shift of spectral weight from low to
high binding energies by $\sim 1$ eV, {\it i.e.} an energy of the
order of the bare bandwidth and hence much larger than the thermal
energy scale. This seems to rule out conventional electron-phonon
coupling as the origin of the temperature dependence, though we
cannot completely exclude the possibility of non-linear coupling
effects.\cite{Gutfreund80}

A much more natural explanation of the observed temperature
effects can be inferred from calculations for the quarter-filled
1D $tJ$ model in the strong-coupling limit ($J/t \rightarrow 0$,
corresponding to the $U/t \rightarrow \infty$ case of the Hubbard
model).\cite{Penc97b} Here it was found that, compared to the zero
temperature spectrum, considerable spectral weight is
redistributed from the ''spin'' peak at the Fermi level to the
bottom of the ''charge'' band at $-2t$ already at temperatures $0
< k_B T << t$, exactly as observed in our data. We are not aware
of similar calculations for moderate interaction strengths, but we
expect this result to hold qualitatively also for finite $J/t$ or
$U/t$, respectively.

In conclusion, both the dispersive behavior of the TCNQ-derived
ARPES structures as well as the temperature dependence of the
spectra are found to be well accounted for, in parts even
quantitatively, by the finite-energy spectral properties of the 1D
Hubbard model. The observed discrepancies to band theory thus
appear as a consequence of spin-charge separation, which occurs in
that model on all energy scales.\cite{Carmelo03a} In this
interpretation our ARPES results on TTF-TCNQ represent the first
spectroscopic observation of spin-charge separation in a quasi-1D
metal on an energy scale of the conduction band width. It is
interesting to note that there exists independent experimental
support for the occurence of spin-charge decoupling in TTF-TCNQ
from the contrasting temperature dependence of conductivity and
spin susceptibility. \cite{Tomkiewicz77}

We close this section with a discussion of the spectral onset near
$E_F$, for which the 1D Hubbard model predicts a low energy behavior
$\propto |E-E_F|^{\alpha}$ with the exponent ranging between $\alpha =
0$ and $\alpha = 1/8$ for $U/t\rightarrow 0$ and
$U/t\rightarrow\infty$, respectively.\cite{Voit95} This is in clear
contrast to our experimental observation of an almost {\it linear}
energy dependence, for which there are various possible explanations.
First of all, the physics of the 1D Hubbard model is expected to be
applicable only for excitation energies larger than the transverse
transfer integrals associated with interchain hopping (the DFT band
dispersions of Fig.~\ref{bandstructure} give an estimate of the
relevant energy scale). In addition, long-range interactions beyond a
simple Hubbard model (e.g.~the effect of nearest neighbor interaction,
which may be non-negligible for TTF-TCNQ \cite{Hubbard78}) are capable
to increase the onset exponent up to $\alpha \sim 1$.
\cite{Zhuravlev01} Unfortunately, the spectral properties of extended
Hubbard models at higher binding energies are not well known. Finally,
it has recently been argued that low-energy power law exponents of the
order of unity can also be caused by impurities and/or defects on the
surface of an organic conductor, which localize the 1D electrons to
strands of finite length, leading to the concept of a "bounded
Luttinger liquid".\cite{Voit00}

However, it seems likely that the failure of the simple Hubbard model
at {\it low} energies is not just a purely electronic effect. Rather,
one should also expect pronounced contributions by electron-phonon
coupling, which after all is strong enough to drive a Peierls
transition at low temperatures. On the other hand, a simple
interpretation of our linear spectral onset in terms of a Peierls
pseudogap due to CDW fluctuations above $T_P = 54$ K must be ruled
out, as the size of the underlying low-temperature Peierls half-gap is
only 20 meV,\cite{Kagoshima88} much smaller than the energy range of
the onset ($\sim 100$ meV). Furthermore, $2 k_F$ CDW fluctuations
disappear already at 150 K, while $4 k_F$ fluctuations - though still
observable at 300 K - strongly weaken with increasing
temperature.\cite{Pouget76} In contrast, the ARPES spectral weight
near $E_F$ is found to become reduced (while still being
linearly energy-dependent) from low to high temperatures
(cf.~Fig.~\ref{T-spectra}). The large energy range of the spectral
onset indicates the importance of coupling to other phonons than those
involved in the Peierls transition and is consistent with the phonon
spectrum of TTF-TCNQ, which indeed reaches up to $\sim 200$
meV.\cite{Eldridge98} Even so, any detailed understanding of the
spectral properties of TTF-TCNQ at low energies will require the
consideration of electronic correlations {\it and} electron-phonon
coupling effects on an equal footing, which remains to be a challenge
to modern solid state theory. Whatever the details of any such
description, our above results indicate that its high-energy physics
must be close to that of the 1D Hubbard model.

\section{Conclusion}

The electronic structure of TTF-TCNQ above the Peierls transition
as probed by ARPES deviates significantly from DFT band
calculations. The experimental observation of an approximate
doubling of the overall conduction band width relative to band
theory is attributed to a structural relaxation of the topmost
molecular layers. When accounted for an enhanced electron hopping
integral at the surface, the spectra of the TCNQ-derived bands can
be brought into consistent and even quantitative agreement with
the theoretical finite-energy single-particle spectrum of the 1D
Hubbard model. This picture is further
supported by a temperature-dependent redistribution of spectral
weight over energies much larger than the thermal energy. Within
this interpretation our experimental results provide spectroscopic
evidence for spin-charge separation on an energy scale of the
conduction band width. In contrast, the spectral behavior at low
binding energies is found to deviate from that of the simple 1D
Hubbard model, possibly due to higher dimensional effects combined
with the additional importance of electron-phonon coupling and
possibly also long-range electron-electron interaction. TTF-TCNQ
thus represents an interesting model system to study electronic
correlation effects in a 1D metal.

\acknowledgments

We gratefully acknowledge S. Hao, C. Janowitz
and G. Reichardt for technical support at BESSY, and F. Gebhard, E.
Jeckelmann, J. M. B.~Lopes dos Santos, A.~Muramatsu, K.~Penc, and
J.~Voit for stimulating discussions. This work was supported by the
DFG (CL 124/3 and SFB 484), the BMBF (Grant 05SB8TSA2), and the FCT
(Ph.D.~Grant BD/3797/94).

\widetext
\break

\begin{figure}
\includegraphics[width=16cm]{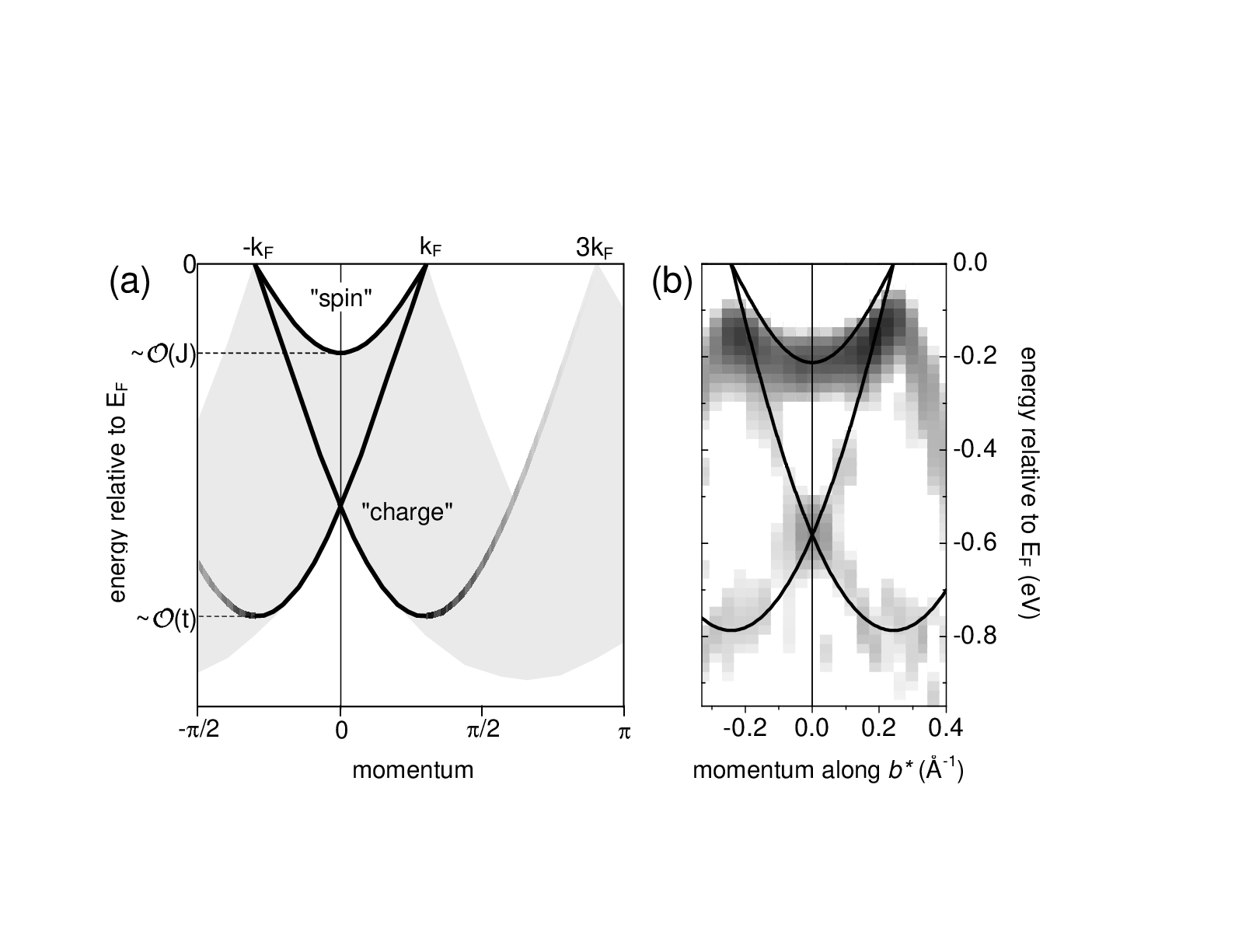}
\caption{(a) Schematic electron removal spectrum of the doped 1D
Hubbard model with band filling $1/2 < n < 2/3$. The shaded region
denotes the continuum resulting from spin-charge separation. The
solid curves indicate the dispersions of the ''spin'' and
''charge'' singularity branches (see text for details). (b)
Theoretical spin and charge branch dispersions of the 1D Hubbard
model calculated for $U = 1.96$ eV, $t = 0.4 $ eV, and $n = 0.59$
in comparison to the ARPES dispersions of the TCNQ-derived
conduction band complex (measured with He I radiation,
representation as in Fig.~\ref{intensity map}).} \label{Hubbard
model spectra}
\end{figure}


\begin{thebibliography}{10}
\bibitem{criticalpoints}
Our Z-point is identical to the Y-point of
Ref.~\onlinecite{Zwick98}.
\bibitem{footnote_normalization}
The spectra have been measured with He I radiation (21.2 eV) from
a discharge
  source. Here the photocurrent above the Fermi edge results from inelastic
  electrons due to photoexcitation by He II (40.8 eV) satellite radiation and
  is proportional to the intensity of the He I main line.
\bibitem{footnote_on_splitting}
In this interpretation the doublet character of the TCNQ bands
predicted by
  band theory is assumed to be unresolved in the ARPES spectra.
\end{thebibliography}
\end{document}